\documentclass[a4paper]{jpconf}

\usepackage[english]{babel}
\usepackage[dvips]{graphicx}
\usepackage{verbatim}
\usepackage{amsfonts}
\usepackage{makeidx}
\usepackage{color}
\usepackage{amsmath}
\usepackage{subfigure}
\usepackage{multirow}
\usepackage{rotating}
\usepackage{iopams}

\usepackage{fancyhdr}
\usepackage{emptypage}
\usepackage[numbers,square,sort&compress,comma]{natbib}

\newcommand{\gev}{\mathrm{GeV}}
\newcommand{\tev}{\mathrm{TeV}}

\newcommand{\Npp}{N_{\rm pp}}
\newcommand{\Naa}{N_{\rm AA}}
\newcommand{\Ncoll}{N_{\rm coll}}
\newcommand{\Raa}{R_{\rm AA}}
\newcommand{\RAA}{R_{\rm AA}}
\newcommand{\TAA}{T_{\rm AA}}

\newcommand{\pt}{p_{\rm T}}

\newcommand{\DtoKpi}{{\rm D^0\to K^-\pi^+}}
\newcommand{\DtoKpipi}{{\rm D^+\to K^-\pi^+\pi^+}}
\newcommand{\DstartoDpi}{{\rm D^{*+}\to D^0\pi^+}}
\newcommand{\Dzero}{{\rm D^0}}
\newcommand{\Dstar}{{\rm D^{*+}}}
\newcommand{\Dplus}{{\rm D^+}}

\newcommand{\vtwo}{v_2}

\begin{document}


%
%
%
%
%
\title{Measurement of heavy-flavor production in Pb--Pb collisions at the LHC with ALICE}
%
\author{Robert Grajcarek for the ALICE collaboration}
\address{Physikalisches Institut der Universit\"{a}t Heidelberg,
Im Neuenheimer Feld 226,\\69120 Heidelberg, Germany}

\ead{grajcarek@physi.uni-heidelberg.de}

%
%
\begin{abstract}
A Large Ion Collider Experiment (ALICE) at the Large Hadron Collider (LHC) has been built in order to study the Quark--Gluon Plasma (QGP) created in high-energy nuclear collisions. As heavy--flavor quarks are produced at the early stage of the collision, they serve as sensitive probes for the QGP. The ALICE detector with its capabilities such as particle identification, secondary vertexing and tracking in a high multiplicity environment can address, among other measurements, the heavy-flavor sector in heavy--ion collisions.\\ 
We present latest results on the measurement of the nuclear modification factor of open heavy-flavors as well as on the measurement of open heavy-flavor azimuthal anisotropy $\vtwo$ in Pb-Pb collisions at $\sqrt{\mathrm{s}_{_{\mathrm{NN}}}} = 2.76$ TeV. \\
Open charmed hadrons are reconstructed in the hadronic decay channels $\DtoKpi$, $\DtoKpipi$, and $\DstartoDpi$ applying a secondary decay-vertex topology. Complementary measurements are performed by detecting electrons (muons) from semi--leptonic decays of open heavy-flavor hadrons in the central (forward) rapidity region.
\end{abstract}
%



\section{Introduction} 
A Large Ion Collider Experiment (ALICE)~\cite{alice} at the Large Hadron Collider (LHC) has been built to investigate the properties of a color-deconfined state of strongly interacting matter, that is formed in ultra-relativistic Pb-Pb collisions. Heavy--flavor quarks (charm and beauty) are well--suited probes to study the hot and dense matter, called Quark--Gluon Plasma (QGP). Due to their large mass heavy--flavor quarks are produced in initial hard scattering processes on a very short time scale of the order of $\tau\approx1/2\textnormal{m}_q$. After production they interact with the formed QGP and because of their early production they are sensitive to the full history of the collision. The experimental measurements to access the heavy-flavor sector, which are covered by these proceedings, are:
\begin{itemize}
	\item Fully reconstructed open charmed mesons $\Dzero$, $\Dplus$ and $\Dstar$ in the following hadronic decay channels:
	\subitem $\bullet~\DtoKpi$ and the charge conjugates.
	\subitem $\bullet~\DtoKpipi$ and the charge conjugates.
	\subitem $\bullet~\DstartoDpi$ and the charge conjugates.
	\item Electrons and muons from charm and beauty semi--leptonic hadron decays.
\end{itemize}
One important and well established quantity to characterize the interaction of heavy quarks with the medium is the nuclear modification factor $\Raa$ studied as a function of transverse momentum $\pt$. This quantity is defined as the ratio of the $\pt$ spectrum measured in AA collisions to the proton--proton spectrum at the same center--of--mass energy $\sqrt{s}$ scaled by the number $\Ncoll$ of binary nucleon--nucleon collisions in the AA collision:
\begin{equation} 
\label{eq:RAA}
\Raa(\pt) = \frac{1}{\langle \Ncoll \rangle} \cdot \frac{{\rm d}\Naa / {\rm d}\pt}{{\rm d}\Npp / {\rm d}\pt}        = \frac{1}{\langle \TAA \rangle} \cdot \frac{{\rm d}\Naa / {\rm d}\pt}{{\rm d} \sigma _{\rm pp} / {\rm d}\pt} 
\end{equation} 
with $\langle \TAA \rangle$ being the average nuclear overlap function calculated in a Glauber model of the AA collision geometry. $\Ncoll$ and $\langle \TAA \rangle$ are dependent on the impact parameter b of the colliding nuclei, i.e.~on the collision centrality. The ratio $\Raa(\pt)$ equals unity in case of no nuclear effects. In a QGP quarks and gluons are expected to interact with the hot medium and lose energy. Due to their smaller color charge than gluons, quarks are expected to lose less energy than gluons. Furthermore, QCD--theory predicts that the ''dead-cone-effect'' reduces the energy loss of massive charm and beauty quarks with respect to light quarks, see~\cite{RAAtheory1,RAAtheory2,RAAtheory3}. Therefore the expected nuclear modification ordering in the range $\pt~\lesssim~10~\gev/c$, where the heavy-quark masses are not negligible with respect to their momenta, is the following: $\Raa^{\rm \pi}<\Raa^{\rm D}<\Raa^{\rm B}$, with $\Raa^{\rm \pi}$ denoting the nuclear modification factor of mostly gluon-originated pions, $\Raa^{\rm D}$ the nuclear modification factor of open charm D mesons and $\Raa^{\rm B}$ the nuclear modification factor of open beauty B mesons, see~\cite{RAAtheory4}.\\
Another important quantity is the azimuthal anisotropy parameter $\vtwo$. In heavy--ion collisions the spatial anisotropy in the overlapping zone in non-central collisions is converted into a momentum anisotropy of the produced particles, if the in--medium mean free path allows for sufficient re--scattering. In this case, the azimuthal distribution
of the emitted particles $\frac{dN}{d\phi}$ reflects the initial anisotropy, which can be parametrized via the coefficients of a Fourier expansion
\begin{equation} 
\label{eq:v21}
\frac{dN}{d\phi} = \frac{N_0}{2\pi}\left\{1+2v_1\cos(\phi-\Psi_{r})+2v_2\cos(2\phi-2\Psi_{r})+...\right\}
\end{equation}
where $\Psi_{r}$ is the azimuth of the reaction plane in each expansion order $n$. The azimuth of the reaction plane for each harmonic $n$ is determined experimentally event-by-event. The Fourier coefficients are given by $v_1$,$v_2$,...,$v_n$. The origin of a non--zero 
$\vtwo$ is expected to come from different pressure gradients in--reaction plane and out--of--reaction plane at low $\pt$, which is known as elliptic flow, and from the difference in the path length which results in different energy loss in--reaction plane and out--of--reaction plane at high $\pt$. The $\vtwo$-extraction method presented in short here is the ''event plane method''. More details about the azimuthal anisotropy parameter $\vtwo$, alternative methods of its determination and physics background can be found in~\cite{flow}. \\
The proceedings are organized as follows. In Sec.~\ref{sec:detector} the ALICE detector and the data used for the heavy-flavor analysis are described. Section~\ref{sec:pp} presents the results of the heavy-flavor measurements in proton--proton collisions. Next, the $\RAA$ results for heavy-flavor are presented in Sec.~\ref{sec:RAA}. Finally, Sec.~\ref{sec:v2} summarizes the ALICE measurement on the azimuthal anisotropy parameter $\vtwo$ of heavy-flavor hadrons.

\section{The ALICE detector} \label{sec:detector}

For the results presented in the next section the signals of the following sub-detectors of ALICE were used:
\begin{itemize}
	\item \textbf{Inner Tracking System:}\\ The \textbf{I}nner \textbf{T}racking \textbf{S}ystem (ITS) is located at midrapidity and at radii between 4~cm and 44~cm around the interaction point and consists of silicon pixel, silicon drift and silicon strip detectors. It provides a high spatial resolution, secondary vertexing capabilities and is used for trigger purposes.
	\item \textbf{Time Projection Chamber:} \\ The \textbf{T}ime \textbf{P}rojection \textbf{C}hamber (TPC) is located at midrapidity at radii between 85 cm and 250 cm around the interaction point and is the main tracking device of the ALICE detector.
	\item \textbf{Transition Radiation Detector:}\\ The \textbf{T}ransition \textbf{R}adiation \textbf{D}etector (TRD) covers the radial range from 290~cm to 370~cm and is located at midrapidity. So far it is used for electron identification at high $\pt$ only in pp collisions for the analyses presented here.
	\item \textbf{Time Of Flight:} \\ The \textbf{T}ime \textbf{O}f \textbf{F}light system is placed at a radial distance of approximately 400~cm at midrapidity and measures the time of flight of particles from the interaction point. The time information is used for particle identification.
	\item \textbf{Muon Spectrometer:}\\ The Muon Spetrometer is placed at forward rapidity ($-4.0 < \eta < -2.5$). A front absorber with a thickness of ten interaction lengths $\lambda_l$ and an additional muon filter of 7 $\lambda_l$ are used to separate muons partially from the background. The remaining background is subtracted offline during data analysis, which is described in Sec.~\ref{sec:pp}.
		\item \textbf{VZERO:} \\ The VZERO detector is composed of two arrays of scintillator tiles covering the full azimuth in the pseudo-rapidity regions $2.8 < \eta < 5.1$ (VZERO-A) and $ -3.7 < \eta < -1.7$ (VZERO-C). It is used for trigger purposes and collision centrality determination.
\end{itemize}
For the results, that are presented in the following, experimental data from the years 2010 and 2011 were used. A minimum--bias trigger using signals in the two scintillator tiles of the VZERO-detector and the silicon pixel detector of the ITS was applied. Moreover, for the Pb--Pb data taken in November 2011, a collision centrality trigger was used in order to enhance the number of events in semi-central (10-50\% most central collisions) and central (0-7.5\% most central collisions) collisions. This trigger was based on the sum of amplitudes in the VZERO-detector. In proton--proton collisions about 100--180 million (depending on the analysis) minimum--bias triggered events were analyzed. In Pb--Pb collisions 17 million minimum--bias triggered events from the 2010 run and 20 million centrality--triggered events (10-50\% most central collisions) from the 2011 run were analyzed.

\section{Heavy-flavor production in pp collisions at $\sqrt{s} = 7~\tev$ and  $\sqrt{s} = 2.76~\tev$} \label{sec:pp}

\begin{figure}[htb]
  \begin{center}
    \includegraphics[width=0.99\textwidth]{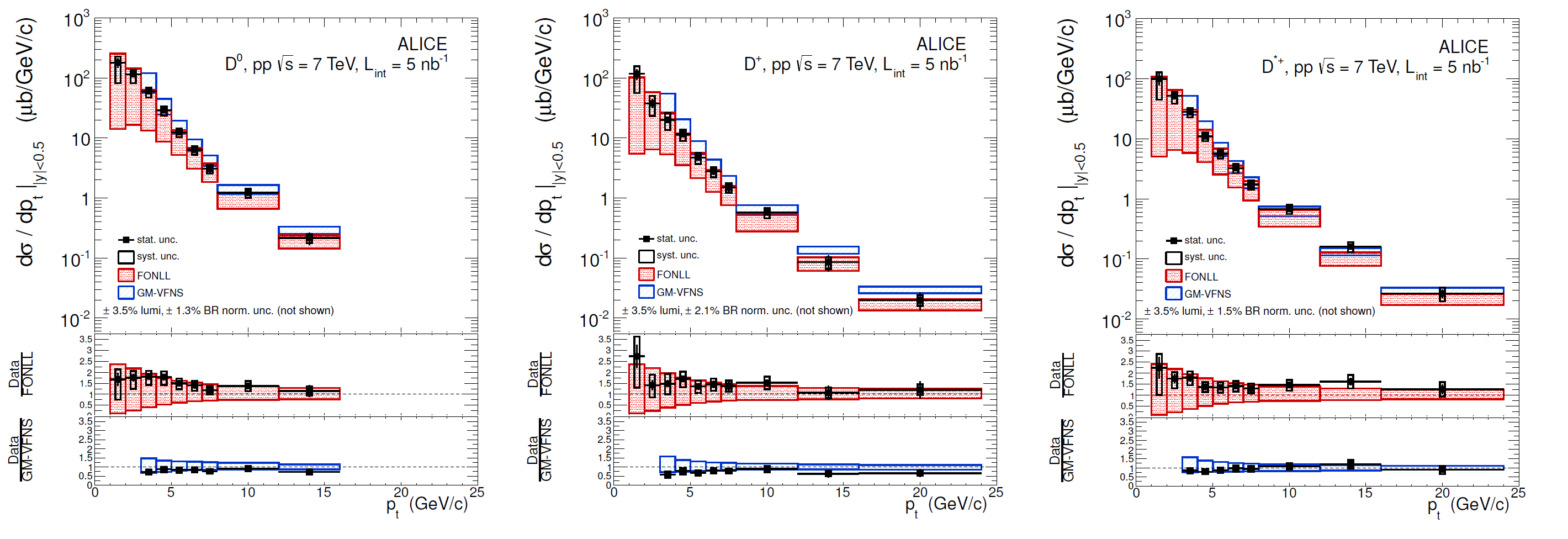}
     \caption{$\pt$-differential production cross sections of $\Dzero$ (left), $\Dplus$ (center) and $\Dstar$ (right) at $\sqrt{s} = 7~\tev$ in $|y| < 0.5$ taken from~\cite{Dmesonspp}, compared to pQCD calculations, namely FONLL~\cite{FONLL,FONLL1,FONLL2} and GM-VFNS~\cite{VFNS1,VFNS2}.}
    \label{fig:Dmesonpp}
  \end{center}
\end{figure}

In order to determine the heavy-flavor nuclear modification factor a cross--section measurement in proton--proton collisions is needed, see Eq.~\ref{eq:RAA}. In this section the analysis details of the heavy--flavor cross--section in proton--proton collisions are presented. \\
Muons from semi-leptonic decays of B mesons and D mesons are measured at forward rapidities with ALICE ($-4.0 < \eta < -2.5$) at $\sqrt{s} = 2.76~\tev$. The raw muon $\pt$ spectrum is corrected for detector efficiency and acceptance using full Monte--Carlo simulations. Moreover, there are two main background sources, which contaminate the heavy-flavor muon spectrum even after online background suppression (for details see~\cite{muons1}), and need to be subtracted:
\begin{itemize}
	\item Muons stemming from primary pion and kaon decays. 
	\item Muons that are produced inside the absorber due to interaction of the primary kaons and pions with the material.
\end{itemize}
Monte--Carlo event generators (PHOJET~\cite{Phojet} and PYTHIA~\cite{Pythia1,Pythia2,Pythia3}) are used to simulate these background sources. The kaon and pion spectra generated according to these generators have been checked against proton--proton data measured by ALICE. For more details see~\cite{muons1}.
\begin{figure}[htb]
  \begin{center}
    \includegraphics[width=0.99\textwidth]{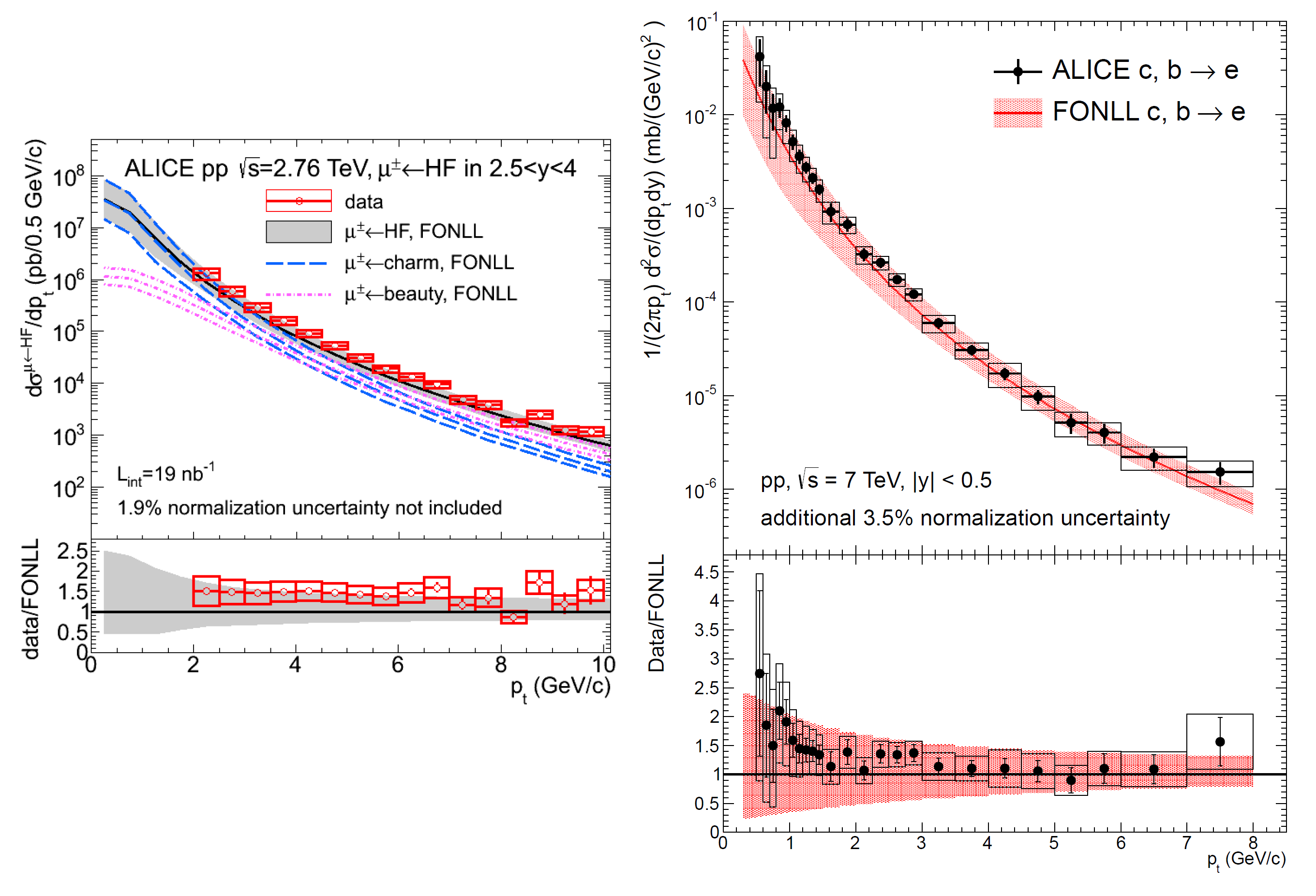}
     \caption{$\pt$-differential production cross sections of heavy-flavor decay muons taken from~\cite{muons1} in $-4.0 < \eta < -2.5$ at $\sqrt{s} = 2.76~\tev$ (left) compared to FONLL calculations~\cite{FONLL}. $\pt$-differential heavy--flavor electron cross section taken from~\cite{electronspp} in $|y| < 0.5$ at $\sqrt{s} = 7~\tev$ (right, upper plot) and the ratio data to FONLL~\cite{FONLL,FONLL1} (right, lower plot).}
    \label{fig:leptspectraPP}
  \end{center}
\end{figure}
The background subtracted $\pt$-differential heavy-flavor muon spectrum $\textnormal{d}\sigma/\textnormal{d}\pt$ taken from~\cite{muons1} is shown in Fig.~\ref{fig:leptspectraPP} (left). The corresponding pQCD prediction~\cite{FONLL} agrees with our data within uncertainties.\\
At midrapidity the heavy-flavor sector is accessed via semi-electronic decays of B mesons and D mesons. The crucial ingredient for this measurement is a clean electron sample. For this purpose the signals of the TPC, TOF and TRD are combined. After exploiting the signals from the three detectors, the remaining hadron contamination is of the order of 10\%. More details about the particle identification procedure are given in~\cite{electronspp}. The main background source for the heavy-flavor electron spectrum are electrons originating from photon conversions in the detector material and in the beam pipe. The origin of these photons are neutral pion decays, with a branching ratio B.R.~($\pi^{0}\rightarrow\gamma\gamma$)~=~98.8\%, abundantly produced in proton--proton collisions at midrapidity. A large fraction of this background source is suppressed by requiring a signal from the electron--candidate track in the innermost layer of the ITS, which is located at a radial distance of only 4~cm from the beam pipe. In order to get the heavy--flavor $\pt$-differential electron spectrum, the raw spectrum after all the selections described above is corrected for efficiency and acceptance using Monte--Carlo simulations with full detector description. Finally, one subtracts the ''cocktail'' of electrons from background sources from the efficiency corrected inclusive electron spectrum. The most contributing components of the cocktail are
\begin{itemize}
	\item conversions of non--direct (e.g. from neutral pions) and direct pQCD photons in the beam pipe and the innermost silicon pixel layer of the ITS.
	\item Dalitz decays of $\pi^{0}$, $\eta$, $\rho$, $\omega$ and $\phi$ mesons.
\end{itemize}
For more details about the cocktail background calculation see~\cite{electronspp}. Figure~\ref{fig:leptspectraPP} (right, upper plot) taken from~\cite{electronspp} shows the $\pt$-differential heavy--flavor electron cross section in $|y| < 0.5$ at $\sqrt{s} = 7~\tev$ and the ratio data to FONLL~\cite{FONLL,FONLL1} (right, lower plot).\\
The third method to access the heavy-flavor sector is the full reconstruction of charm mesons at midrapidity in the hadronic decay channels. The decays $\DtoKpi$, $\DtoKpipi$ and $\DstartoDpi$ are identified via their reconstructed invariant mass. A typical signature of a charm meson decay is the presence of two tracks with opposite charges and with a minimum distance of approach to the primary vertex not compatible with zero. Once the secondary vertex is reconstructed, cut variables can be 
calculated to reduce the combinatorial background. One of the most powerful cuts is the cut on the pointing angle, which is the angle between the reconstructed D meson momentum vector and the vector connecting the primary and secondary vertex. Another important cut parameter is the decay length, i.e. the distance between the primary and secondary vertices. For D mesons this distance is in the order of several $100~\mu\textnormal{m}$. Further background rejection is achieved by identifying pions and kaons using combined signals, which the D meson daughters induce in the TPC and the TOF detectors. The raw yields are corrected for efficiency using Monte--Carlo simulation with full detector description. Moreover, the efficiency--corrected D meson spectra are corrected for B meson feed-down. This B--feed--down correction is needed because according to FONLL (fixed-order next-to-leading log)~\cite{FONLL} about 15\% of the D mesons originate from B meson decays. The B meson feed--down correction is performed using the beauty production cross section predicted by FONLL calculation~\cite{FONLL} and full detector simulation. More details about the D meson measurements in proton--proton collisions with ALICE can be found in~\cite{Dmesonspp}. The $\Dzero$, $\Dplus$ and $\Dstar$ $\pt$-differential production cross sections taken from~\cite{Dmesonspp} in $|y| < 0.5$ are shown in Fig.~\ref{fig:Dmesonpp}. Theoretical predictions based on pQCD calculations, namely FONLL~\cite{FONLL, FONLL1, FONLL2} and GM-VFNS~\cite{VFNS1, VFNS2} are in agreement with the data.\\In order to use the measured spectra of fully reconstructed D mesons or electrons originating from semi--leptonic 
D meson and B meson decays at $\sqrt{s} = 7~\tev$ as references for the corresponding $\Raa$ measurement, they need to be scaled down from $\sqrt{s} = 7~\tev$ to $\sqrt{s} = 2.76~\tev$. The scaling factors were obtained as the ratio
of the corresponding spectra provided from the pQCD FONLL calculations at $\sqrt{s} = 2.76~\tev$ divided by those at $\sqrt{s} = 7~\tev$. For more details about the scaling procedure see~\cite{scaling}. The $\pt$-differential cross sections of fully reconstructed D mesons were also obtained at the reference center--of--mass energy of $\sqrt{s} = 2.76~\tev$, but are with limited statistics. It was found that this measurement is consistent with the scaled reference at $\sqrt{s} = 7~\tev$. For details see~\cite{lowenergy}.
\section{The heavy-flavor nuclear modification factor $\RAA$} \label{sec:RAA}
\begin{figure}[htb]
  \begin{center}
    \includegraphics[width=0.99\textwidth]{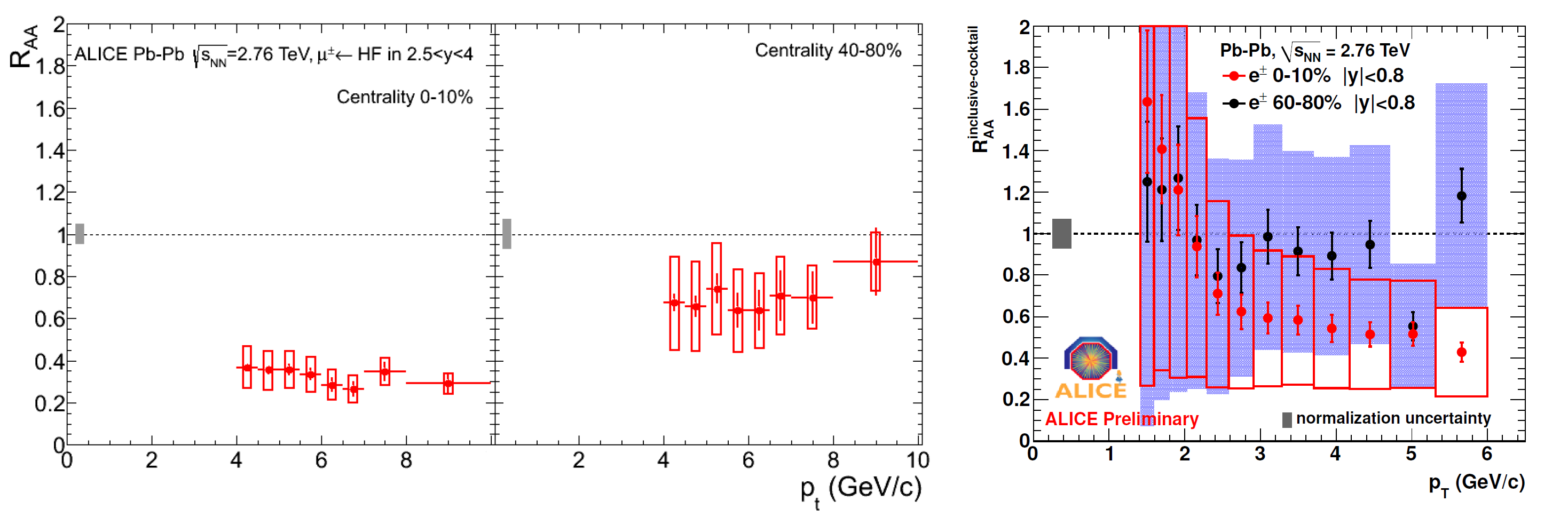}
     \caption{(left) $\Raa$ for heavy--flavor decay muons at forward rapidity in central (0--10\% most central collisions) and peripheral (40--80\% most central collisions) Pb-Pb collisions taken from~\cite{muons1}. (right) Nuclear modification factors for background--subtracted electrons at central rapidity and two different centrality classes, namely 0--10\% and 60--80\% most central. Statistical (bars) and systematic (boxes) uncertainties are shown.}
    \label{fig:RAAleptons}
  \end{center}
\end{figure}

\begin{figure}[htb]
  \begin{center}
    \includegraphics[width=0.99\textwidth]{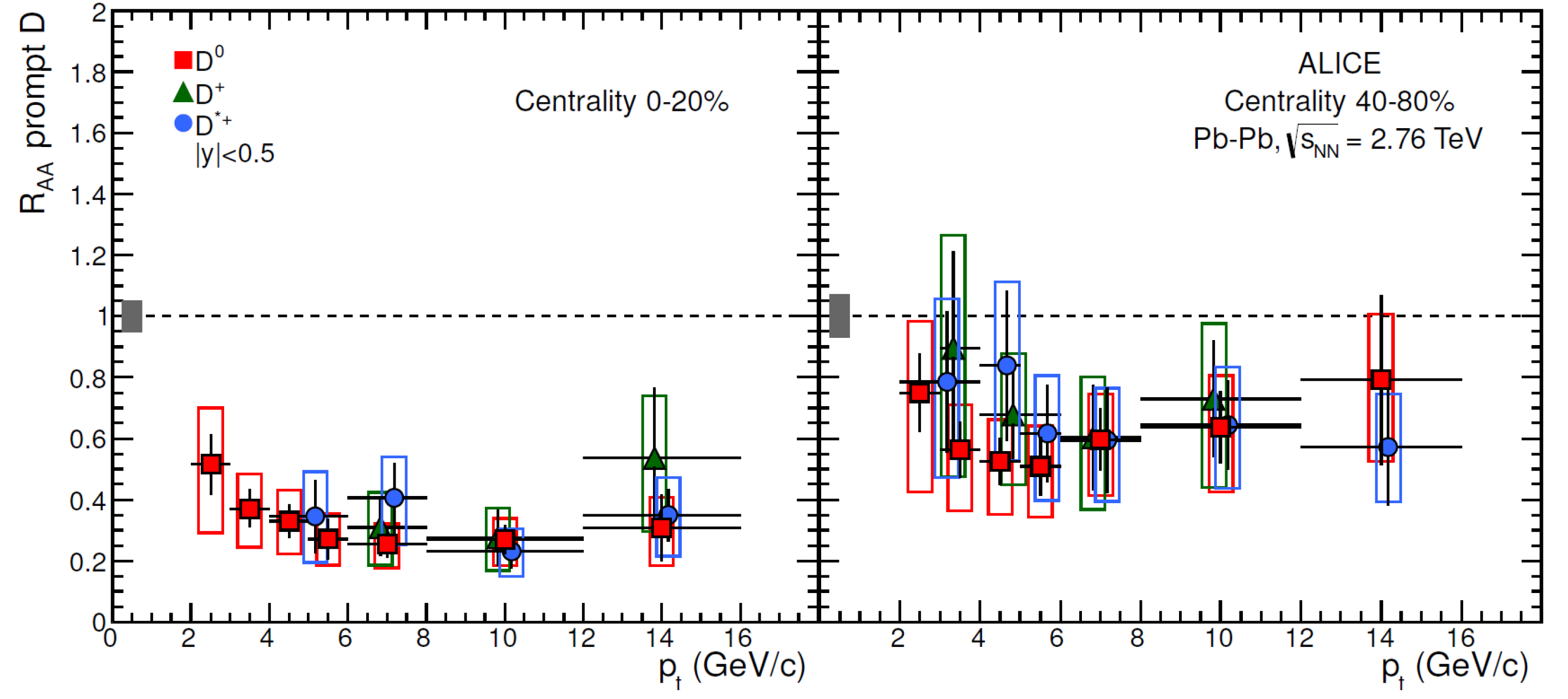}
     \caption{Nuclear modification factors of $\Dzero$, $\Dplus$ and $\Dstar$ mesons for two different centrality classes, namely 0--20\% (left) and 40--80\% (right) most central collisions taken from~\cite{DRAA}. Statistical (bars) and systematic (empty boxes) uncertainties are shown.}
    \label{fig:DRAA}
  \end{center}
\end{figure}

\begin{figure}[htb]
  \begin{center}
    \includegraphics[width=0.99\textwidth]{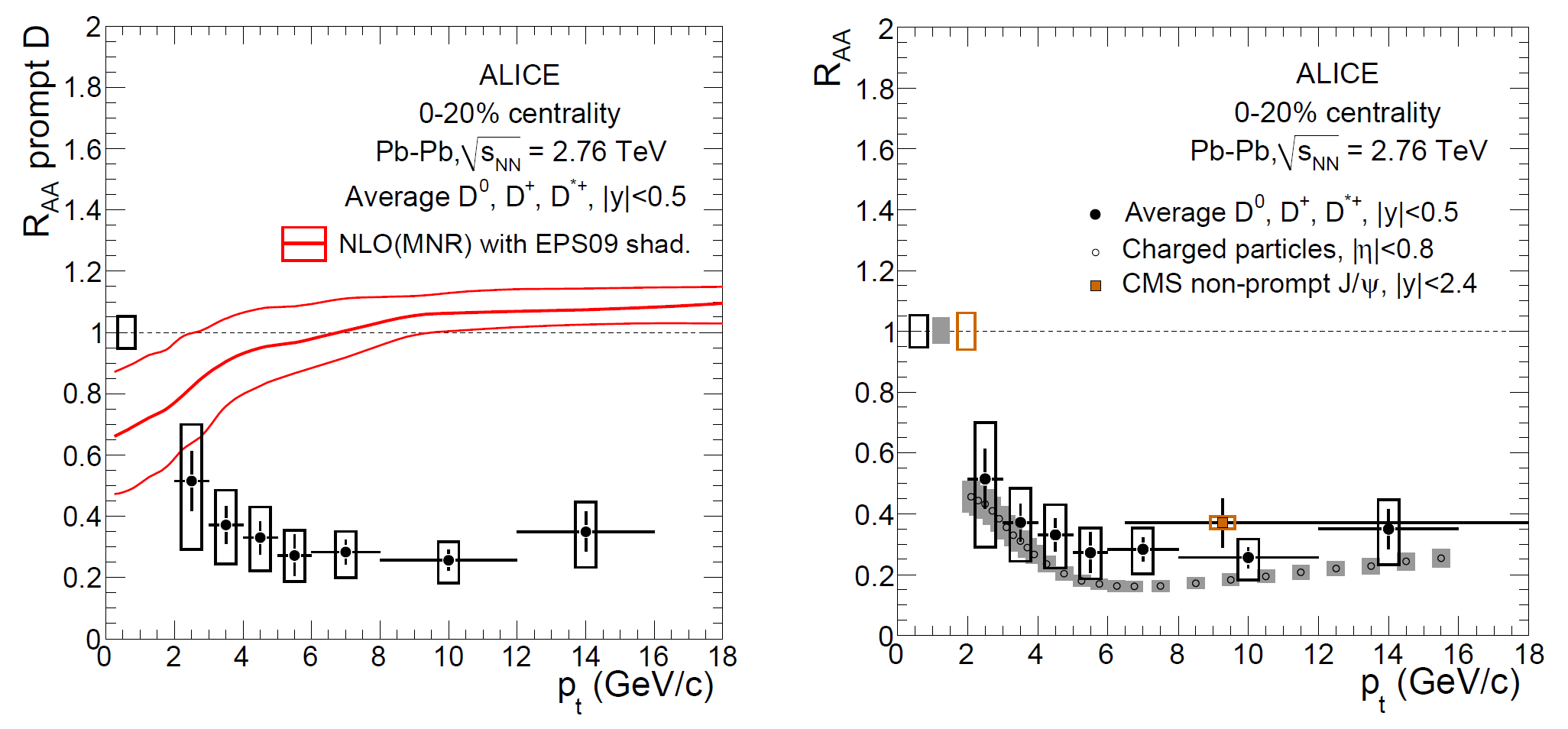}
     \caption{Average nuclear modification factor of D mesons in central Pb--Pb collisions (0--20\% most central collisions) compared to next-to-leading order (NLO)~\cite{RAAmodel1} calculations with CTEQ6M parton distribution functions~\cite{RAAmodel2} including nuclear modified PDF parametrized by EPS09NLO~\cite{RAAmodel3} (left) as well as compared to nuclear modification factors of charged particles~\cite{chargedhadrons} and of $J/\Psi$-mesons stemming from B meson decays measured by the CMS--collaboration~\cite{CMS} (right). Both figures have been taken from~\cite{DRAA}.}
    \label{fig:DRAAmodels}
  \end{center}
\end{figure}

The selection and background subtraction procedures for the single muon and electron heavy--flavor spectrum are the same for Pb--Pb collision data except for minor details described below.\\In case of the muon heavy--flavor spectrum the Pb--Pb analysis differs from the  proton--proton analysis in treatment of the background subtraction. The spectra of the
charged pions and kaons used for background subtraction are taken from simulation (see Sec.~\ref{sec:pp}), while in the Pb--Pb collisions the ones measured by ALICE at midrapidity are utilized. Next, they are extrapolated to forward rapidities using the rapidity shape from event generators and $\Raa$ information based on ATLAS measurements~\cite{muons1}. The $\Raa$ for kaons and pions is assumed to be equal at midrapidity and forward rapidity but is varied within $\pm~100\%$ to evaluate a systematic uncertainty for the $\Raa$ of heavy--flavor decay muons. The nuclear modification factor for the heavy--flavor decay muons taken from~\cite{muons1} is presented in Fig.~\ref{fig:RAAleptons} (left) for two different centrality classes, namely 0--10\% and 40--80\% most central collisions. A suppression factor of about 3 is observed for 0--10\% most central collisions and is significantly smaller in more peripheral (40-80\% most central collisions) collisions. No $\pt$-dependence of the $\Raa$ is observed within uncertainties.\\In case of the electron heavy--flavor spectra the difference in the Pb--Pb analysis with respect to the proton--proton analysis is that the main input to the cocktail ingredient is not the measured neutral pion spectrum but the charged pion spectrum. Moreover, the analysis is performed without the TRD and the measurement is restricted to $\pt < 6~\gev/c$ to keep the hadron contamination below 15\%. The nuclear modification factor for the cocktail--subtracted electrons is presented in Fig.~\ref{fig:RAAleptons} (right) for two different centrality classes, namely 0--10\% and 60--80\% most central collisions. The large systematic uncertainty is dominated by the electron identification uncertainty. A suppression of a factor 1.5--4 is observed for $\pt > 4~\gev/c$ in the 0--10\% most central collisions. In more peripheral collisions (60-80\% most 
central collisions) the suppression factor tends to go to unity. \\
In case of the fully reconstructed open charm D meson analysis in Pb--Pb collisions also the same selection strategy is applied as for the proton--proton case, see Sec.~\ref{sec:pp}. The difference between the two analyses is that the topological cuts are chosen in general tighter to deal with the larger combinatorial background in the Pb--Pb environment. Moreover, in the Pb--Pb case,  a new systematic error source is added to the B meson feed--down correction which is not present in the proton--proton case. This systematic source comes from the fact that the ratio $\Raa^{\rm B}/\RAA^{\rm D}$, which is needed for the B meson feed--down correction, is unknown. To estimate the systematic uncertainty it is varied in the range $0.3 < \Raa^{\rm B}/\RAA^{\rm D} < 3$. This assumed range represents the spread of models predictions for the charm and beauty $\Raa$~\cite{RAAtheory3,RAAtheory4} and results on the $\Raa$ for $J/\Psi$-mesons stemming from B meson decays measured by the CMS collaboration~\cite{CMS}. This variation yields a systematic uncertainty in the final $\RAA^{\rm D}$ of up to 30\% at high $\pt$. For more details about the analysis in Pb--Pb, see~\cite{DRAA}. The nuclear modification factor of prompt $\Dzero$, $\Dplus$ and $\Dstar$ mesons in central (0-20\% most central collisions) and peripheral (40-80\% most central collisions) Pb--Pb collisions taken from~\cite{DRAA} is shown in Fig.~\ref{fig:DRAA}.
A strong suppression up to a factor of 3-4 for $\pt > 5~\gev/c$ for all three meson species is observed in central collisions. At lower $\pt$ and especially in peripheral collisions the suppression gets significantly weaker. Figure~\ref{fig:DRAAmodels} (left) taken from~\cite{DRAA} shows a comparison of the $\RAA$ averaged over the three meson species and a theoretical  
parametrization of the D meson $\RAA$, which was calculated using a pQCD next-to-leading order (NLO)~\cite{RAAmodel1} approach with CTEQ6M parton distribution functions~\cite{RAAmodel2} including only nuclear modified PDF parametrized by EPS09NLO~\cite{RAAmodel3} but no heavy-quark energy loss. For $\pt > 6~\gev/c$ the effect on the $\Raa$ caused by shadowing is only $\pm15\%$ indicating that the strong suppression of D mesons is a final-state effect. The analysis of the p--Pb collisions data planned to be taken at the beginning of 2013 will enable us to disentangle initial--state and final-state effects. Figure~\ref{fig:DRAAmodels} (right) taken from~\cite{DRAA} shows a comparison between the $\RAA$ averaged over the three meson species, the $\RAA$ of charged particles~\cite{chargedhadrons} and the $\RAA$ of $J/\Psi$--mesons from B meson decays measured by the CMS collaboration~\cite{CMS}. There is an indication for $\Raa^{\textnormal{charged}}<\Raa^{\rm D}$. However, the indication is not significant enough with the current uncertainties. The B meson $\Raa$ measured by the CMS collaboration is clearly larger than the $\Raa^{\textnormal{charged}}$ but due to the large error bars and the coarse binning in $\pt$ no conclusion can be drawn for the moment from the comparison of $\Raa^{\rm D}$ with $\Raa^{\rm B}$.
\section{The heavy-flavor azimuthal anisotropy parameter $\vtwo$} \label{sec:v2}
\begin{figure}[htb]
  \begin{center}
    \includegraphics[width=0.99\textwidth]{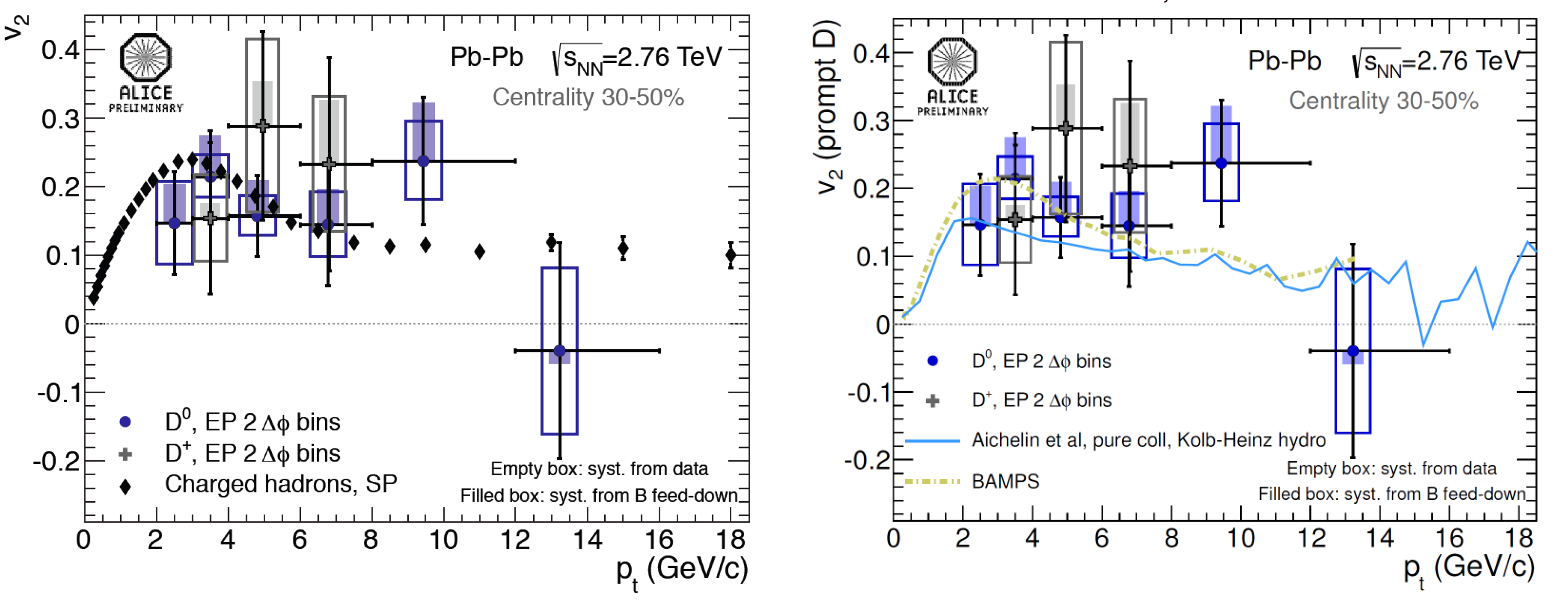}
     \caption{The azimuthal anisotropy parameter $\vtwo$ of D mesons compared with the one of charged particles measured by ALICE~\cite{chargedflow} (left) and theoretical model calculations, BAMPS~\cite{BAMPS} and Aichelin et al.~\cite{Aichelin} (right). Statistical (bars), systematic (empty boxes) uncertainties and systematic uncertainties from B--feed--down correction (shaded boxes) are shown.}
    \label{fig:v2D}
  \end{center}
\end{figure}

In this section the $\pt$--differential azimuthal anisotropy parameter $\vtwo$ is presented for fully reconstructed $\Dzero$ and $\Dplus$ mesons. In order to extract the $\pt$--differential D meson $\vtwo$, the ''event plane method''~\cite{flow} is applied. Due to limited statistics the D meson yield is extracted only in two angular regions with respect to the estimated event plane for every event and in several $\pt$-bins. The azimuth of the event plane is determined event by event exploiting the azimuthal anisotropy of tracks (see~\cite{flow}) measured by the TPC only. One can show using Eq.~\ref{eq:v21} that the observed D meson $\vtwo$ is given by
\begin{equation} \label{eq:v23}
\vtwo\left(\pt\right) = \textnormal{R}_2\vtwo^{obs}\left(\pt\right)= \textnormal{R}_2\frac{\pi}{4}\frac{\textnormal{N}_\textnormal{in-plane}\left(\pt\right)-\textnormal{N}_\textnormal{out-of-plane}\left(\pt\right)}{\textnormal{N}_\textnormal{in-plane}\left(\pt\right)+\textnormal{N}_\textnormal{out-of-plane}\left(\pt\right)}
\end{equation}
with $\textnormal{N}_\textnormal{in-plane}$ being the D meson yield extracted in the event plane and $\textnormal{N}_\textnormal{out-of-plane}$ the D meson yield extracted perpendicular to the event plane. Moreover, a correction factor $\textnormal{R}_2$ accounting for the resolution of the event plane~\cite{flow} is applied. Figure~\ref{fig:v2D} (left) shows the $\pt$-differential $\Dzero$--$\vtwo$ and 
$\Dplus$--$\vtwo$ compared with the charged particle $\vtwo$ measured by ALICE in the same rapidity region~\cite{chargedflow}. Within the large uncertainties the D meson $\vtwo$ is consistent with the $\vtwo$ of charged particles. Figure~\ref{fig:v2D} (right) shows the comparison of D meson $\vtwo$ with calculations using two different theoretical models, BAMPS~\cite{BAMPS} and Aichelin et al.~\cite{Aichelin}. The model BAMPS~\cite{BAMPS} is a partonic transport model computed with a Boltzmann approach to multi-parton scattering. The full space-time evolution of the QGP is obtained by solving the Boltzmann equation for on--shell partons and pQCD interactions. The model introduced in~\cite{Aichelin} is based on a collisional energy loss mechanism for heavy quarks. Both models describe the $\pt$-dependence of the D meson $\vtwo$ properly within the large experimental uncertainties. 
\section{Conclusion} \label{sec:concl}
ALICE has measured the heavy-flavor nuclear modification factor $\Raa$ in Pb--Pb collisions via full reconstruction of open charm mesons and via semi-leptonic decays of charm and beauty hadrons. Both in the reconstruction analysis of open charm mesons and in the semi-leptonic decay channel of open heavy-flavor mesons a strong suppression comparable to the suppression of charged particles up to a factor of 4 is seen in the in central Pb--Pb collisions. A hint of $\Raa^{\textnormal{charged}}<\Raa^{\rm D}$  is observed in the data. The $\Raa$ analysis of the high--luminosity run from November 2011 will enable ALICE to precisely measure the heavy--flavor quark $\Raa$. A first analysis of the 2011 run focusing on the $\pt$-differential azimuthal anisotropy parameter $\vtwo$ reveals a preliminary result of the $\vtwo$ of open charm hadrons. Within the large uncertainties the $\vtwo$ results are properly described by two theoretical models. Finally, a planned p--Pb run at the beginning of 2013 will enable ALICE to quantify initial-state nuclear effects.

\providecommand{\newblock}{}


\end{document}